\documentclass[reprint, amsmath, amssymb, aps, prb, superscriptaddress, nofootinbib]{revtex4-2}

\usepackage[dvipdfmx]{graphicx,color}
\usepackage{dcolumn}
\usepackage{bm}
\usepackage{hyperref}
\usepackage{mathrsfs}
\usepackage{physics}
\usepackage{xcolor}
\usepackage{txfonts}

\DeclareMathOperator{\e}{e}

\usepackage{ulem} 
\usepackage{color}

\usepackage{CJKutf8}

\begin{document}

\preprint{Lattice}

\title{Multi-tunneling effect of nonreciprocal Landau-Zener tunneling: Insights from DC field responses}

\author{Ibuki Terada}
\affiliation{
Department of Physical Science, Ritsumeikan University, Shiga 525-8577, Japan
}
\author{Sota Kitamura}
\affiliation{
Department of Applied Physics, The University of Tokyo, Hongo, Tokyo, 113-8656, Japan
}
\author{Hiroshi Watanabe}
\affiliation{
Research Organization of Science and Technology, Ritsumeikan University, Shiga 525-8577, Japan
}
\author{Hiroaki Ikeda}
\affiliation{
Department of Physical Science, Ritsumeikan University, Shiga 525-8577, Japan
}

\date{\today}

\begin{abstract}
Recent advancements in laser technology have spurred growing interest in nonlinear and nonequilibrium phenomena. Here, we investigate the geometric aspects of quantum tunneling and the nonreciprocal response, particularly focusing on the shift vector, in noncentrosymmetric insulators under a strong DC electric field. In insulators under a strong electric field, electrons undergoing Bloch oscillations interfere with each other by passing through different paths via Landau-Zener tunneling. We found that the interference effect due to multi-tunneling causes the oscillating nonreciprocal response that is significantly amplified with increasing electric field intensity. We also clarified the role of the shift vector in the interference conditions through an analysis of the nonequilibrium steady state. These results will contribute significantly to advancing a systematic understanding of quantum geometric effects in the nonperturbative regime.
\end{abstract}

\maketitle

\section{introduction}
Advances in laser technology have driven active studies of nonequilibrium phenomena occurring under strong electric fields, such as Floquet engineering~\cite{annurev:/content/journals/10.1146/} and nonlinear responses.
Landau-Zener tunneling~\cite{Landau1,Landau2,Zener,Stueckelberg,Majorana,IVAKHNENKO20231}, a key nonperturbative process describing electron transitions between energy bands, is crucial in strong electric fields and central to explaining nonlinear phenomena like photocurrent~\cite{Nature.7675.224,10.21468/SciPostPhys.11.4.075}  and high harmonic generation~\cite{NatPhoto.2.119,NatPhys.2.138}.
On the other hand, in recent years, geometrical aspects of Bloch electrons have become increasingly important in condensed matter physics, and studies of novel quantum phases in topological insulators~\cite{RevModPhys.82.3045,RevModPhys.83.1057} and Dirac/Weyl semimetals~\cite{RevModPhys.90.015001,RevModPhys.93.025002} have progressed rapidly. Topological response phenomena such as the quantum Hall effect~\cite{RevModPhys.82.1539,annurev:/content/journals/10.1146/annurev-conmatphys-031115-011417,RevModPhys.95.011002}, topological magnetoelectric effect~\cite{PhysRevB.78.195424,PhysRevLett.102.146805} are also of great interest.

In this context, geometric effects in quantum tunneling processes have also attracted much attention, including the study of the nonreciprocal nature of tunneling probabilities~\cite{doi:10.1098/rspa.1990.0096,PhysRevA.49.R2217,CommunPhys.1.63}, modification of adiabatic conditions~\cite{PhysRevA.77.062114,PhysRevA.97.032124}, and counter-diabatic driving~\cite{JPhysChemA.46.9937,Berry_2009,PhysRevLett.111.100502}. 
An important geometric quantity in these phenomena is the shift vector, which characterizes how the center of the electron cloud shifts during electron transitions in noncentrosymmetric systems~\cite{CommunPhys.1.63,10.21468/SciPostPhys.11.4.075}.
The shift vector is widely recognized as a key element in the mechanism of the shift current, a type of geometrical photocurrent~\cite{PhysRevLett.109.236601,npjComputMater.1.16026,doi:10.1126/sciadv.1501524,doi:10.7566/JPSJ.92.072001}. 
In quantum tunneling processes, the shift vector changes the effective thickness of the tunnel barrier depending on the direction of the electric field, offering an intuitive and physical explanation for the nonreciprocal Landau-Zener tunneling~\cite{CommunPhys.1.63}.

While nonreciprocal current responses and photocurrent generation via circularly polarized light have been proposed~\cite{10.21468/SciPostPhys.11.4.075,PhysRevB.102.245141}, these studies have primarily focused on single tunneling events, without considering the effects of periodic tunneling events. 
In lattice systems, electrons traveling along different paths interfere with each other due to optical transitions and quantum tunneling.
The significance of such interference effects has been confirmed in DC current generation under AC electric fields, irrespective of the electric field strength~\cite{PhysRevLett.78.306, NanoLett.4.1293,Nature.7675.224}. 

In a lattice system under a DC electric field, Bloch oscillations interfere with each other via Landau-Zener tunnelings, resulting in Bloch-Zener oscillations~\cite{Breid_2006,Breid_2007}. Although this phenomenon is challenging to observe in materials, 
it has recently been observed in binary superlattices~\cite{PhysRevLett.102.076802}, along with photonic crystals~\cite{PhysRevLett.96.053903,PhysRevLett.121.033904} and cold atoms in optical lattices~\cite{PhysRevLett.76.4508}. 
Notably in zero-gap semiconductors like graphene, Bloch-Zener oscillations give rise to intriguing phenomena, such as negative differential conductance~\cite{PhysRevB.85.115433}. However, the geometric aspects of this interference effect have not yet been fully elucidated.

Our goal is to systematically explore the geometric aspects of multiple tunneling processes in
Bloch-Zener oscillations of noncentrosymmetric materials.
We derive the conditions for constructive interference (CCI) 
and demonstrate that the shift vector not only governs the nonreciprocity of tunneling probabilities but also influences the electric field strength that enhances carrier occupation. The nonreciprocity of the CCI results in a more pronounced nonreciprocal current response compared to single tunneling events, enabling directional control of electron transport through tuning the electric field strength.

This paper is organized as follows. First, in Sec.~\ref{Sec:Formalism}, we formulate a single electron dynamics for two-band lattice systems in the nonequilibrium steady state. Additionally, we demonstrate the effects of the Bloch-Zener oscillation on current responses.  In Sec.~\ref{Sec:GeometricEffects}, we derive the formula for the distribution involving the double-tunneling processes and discuss the geometric effect of multi-tunneling processes on the current response for quantum systems without inversion symmetry. Finally, we discuss candidate quantum systems for the result of this paper and conclude the paper in Sec.~\ref{Sec:Conclusion}. In this paper, we set the physical constants $\hbar=e=1$.

\section{formalism}\label{Sec:Formalism}
\subsection{Quantum kinetic equation}
We investigate a single electron dynamics in gapped lattice systems using the quantum kinetic equation. 
Here, we consider a two-band system with a finite energy gap $\Delta_k=\varepsilon_{k+}-\varepsilon_{k-}>0$, for simplicity.
A DC electric field $E$ is introduced using the Peierls substitution, where the Hamiltonian $\mathcal H_k$ is replaced by $\mathcal H_{k-E(t-t_0)}=\mathcal H_k(t)$, with $t_0$ being the initial time.

The reduced density matrix $\rho_k(t)$ is governed by the $k$-dependent quantum kinetic equation:
\begin{equation}\label{QME}
	\frac{d\rho_k(t)}{dt}=-i[\mathcal H_k(t), \rho_k(t)]+\mathcal D_k(\rho_k(t)),
\end{equation}
where $\mathcal D_k(t)$ represents the dissipation term. 
We here introduce the adiabatic basis~\cite{Comment1},
\begin{align}\label{Snapshotbasis}
	&\ket{\Phi_{k\alpha}(t)}=\e^{-i\theta_{k\alpha}(t)}\ket{u_{k\alpha}(t)},\\
	&\theta_{k\alpha}(t)=\int_{t_0}^t[\varepsilon_{k\alpha}(t')+EA_{k,\alpha\alpha}(t')]dt',
\end{align}
where $\varepsilon_{k\alpha}(t)=\varepsilon_{k-E(t-t_0),\alpha}$ and $\ket{u_{k\alpha}(t)}=\ket{u_{k-E(t-t_0),\alpha}}$ are the snapshot eigenenergies and eigenstates of $\mathcal H_k(t)$, respectively. 
The phase factors $-i\int_{t_0}^t\varepsilon_{k\alpha}(t')dt'$ and $-i\int_{t_0}^tEA_{k,\alpha\alpha}(t')dt'$ are the dynamical phase and the Berry phase with Berry connection $A_{k,\alpha\beta}(t)=i\braket{u_{k\alpha}(t)}{\partial_k u_{k\beta}(t)}$, respectively. 
In this basis, the Hamiltonian in Eq. \eqref{QME} is transformed into the Hamiltonian in the adiabatic basis
\begin{align}\label{EffectiveHam}
&\mathcal W_k(t)=
\left(\begin{matrix}
0&W_k(t)\\
W^*_k(t)&0
\end{matrix}\right),\\
&W_k(t)=E|A_{k,+-}(t)|\e^{i\int_{t_0}^t[\Delta_k(t')+ER_k(t')]dt'+i\arg A_{k,+-}(t_0)},
\end{align}
where $W_k(t)$ and $R_k=A_{k,++}-A_{k,--}-\partial_k\arg A_{k,+-}$ are the transition dipole moment in the adiabatic basis and the shift vector, respectively.

The dissipation term is often determined with the relaxation time approximation (RTA).
However, as demonstrated in Ref.~\cite{PhysRevB.109.L180302}, this approximation fails to accurately represent the conductivity of an insulator. 
The DC conductivity obtained from the RTA shows unphysical linear conductivity for insulators in the linear response regime due to the inadequate treatment of the density matrix.
In Ref.~\cite{PhysRevB.109.L180302}, we addressed this issue and proposed the dynamical phase approximation (DPA) as an alternative to the RTA, which semi-quantitatively reproduces the results of the exact solution across the entire electric field region.
Furthermore, it has been successfully extended to nonlinear response in Ref.~\cite{ICPS2024}.
Based on this DPA formalism, the dissipation term in particle-hole symmetric insulating systems is given as
\begin{subequations}
\begin{align}
&\mathcal D_k=\mathcal D^{(0)}_k+\mathcal D^{(1)}_{k}+\mathcal D^{(2)}_{k}, \\
&[\mathcal D^{(0)}_{k}]_{\alpha\beta}=-\frac{[\rho_k(t)]_{\alpha\beta}-f_D(\varepsilon_{k\alpha}(t))\delta_{\alpha\beta}}{\tau},\\
\label{D_odia}
&[\mathcal D^{(1)}_{k}]_{\alpha\beta}=-\frac{[\mathcal W_k(t)]_{\alpha\beta}\delta f_k(t)}{\tau\Delta_k(t)}\left(1-\frac{Ez^\alpha_k(t)}{\Delta_k(t)}\right),\\
\label{D_dia}
&[\mathcal D^{(2)}_{k}]_{\alpha\beta}=\frac{\alpha|W_k(t)|^2\delta f_k(t)}{\tau\Delta^2_k(t)}\left(1-\frac{Ez^+_k(t)}{\Delta_k(t)}\right)\left(1-\frac{Ez^-_k(t)}{\Delta_k(t)}\right)\delta_{\alpha\beta},
\end{align}
\end{subequations}
where $\tau^{-1}$ denotes the damping rate, $z^\pm_k=R_k\pm i\partial_k\ln|A_{k,+-}|$ and $\delta f_k=f_D(\varepsilon_{k-})-f_D(\varepsilon_{k+})$ 
with $f_D$ being the Fermi-Dirac distribution function.
$\mathcal D^{(0)}_{k}$ is the RTA term, while $\mathcal D^{(1)}_{k}$ and $\mathcal D^{(2)}_{k}$ are the field-induced corrections describing the relaxation from an excited state to a polarized state.

The solution of Eq.~(\ref{QME}) in the adiabatic basis is formally written as
\begin{equation}\label{DM}
\begin{split}
\rho_k(t)&=U(t)\rho_k(t_0)U^\dag(t)\e^{-(t-t_0)/\tau}\\
&\qquad+U(t)\bigg(\int_{t_0}^t U^\dag(s)\Sigma(s)U(s)\e^{-(t-s)/\tau}~ds\bigg)U^\dag(t),
\end{split}
\end{equation}
where $U(t)=U(t,t_0)$ is the time evolution operator of the isolated system,
\begin{equation}
U(t_1,t_2)=\mathcal T \exp\left[-i\int_{t_2}^{t_1}\mathcal W_k(s)ds\right]
\end{equation} and $\Sigma(t)=\mathcal D_k(t)+\rho_k(t)/\tau$.
Note that the density matrix~(\ref{DM}) includes nonperturbative effects via $U(t)$, such as the contribution of the Landau-Zener tunneling. 
Hereafter, we set the initial time $t_0=0$.

\subsection{Dissipative dynamics in lattice systems}\label{DissipativeDynamics}
Let us consider the nonequilibrium steady state of Bloch-Zener oscillations~\cite{Breid_2006,Breid_2007} that arise when a DC electric field is applied to lattice systems. The density matrix describing the nonequilibrium steady state, $\rho^{\rm NESS}_K$, is obtained as the long-time limit in Eq.~\eqref{DM}.
Reflecting the Bloch oscillation, the snapshot eigenenergy and eigenstates are time-periodic functions, $\varepsilon_{k\alpha}(t+T_{\!B})=\varepsilon_{k\alpha}(t)$ and $\ket{u_{k\alpha}(t+T_{\!B})}=\ket{u_{k\alpha}(t)}$, respectively.
Here, $T_{\!B}=2\pi/|E|a_0$ is the period of the Bloch oscillation and $a_0$ is the lattice constant. 
On the other hand, the adiabatic basis, which includes the change in the dynamical phase, satisfies
$\ket{\Phi_{k\alpha}(t+T_{\!B})}=\e^{-i\theta_{k\alpha}(T_{\!B})}\ket{\Phi_{k\alpha}(t)}$.
Then, the Hamiltonian in the adiabatic basis follows the relation below,
\begin{equation}\label{Heff_periodic}
\mathcal W_k(t+T_{\!B})=\mathcal X^{\,-1}_{\!B}\,\mathcal W_k(t)\,\mathcal X_{\!B},
\end{equation}
where $[\mathcal X_{\!B}]_{\alpha\beta}=\e^{-i\alpha\Theta_B/2}\delta_{\alpha\beta}$ and 
\begin{align}
\Theta_B(E)=\int_{\,0}^{\,T_{\!B}}[\Delta_k(t')+ER_k(t')]\,dt'.
\end{align}
Note that $\Theta_B(E)$ is a gauge invariant quantity. 
Consequently, we obtain the following relation for the time evolution $U(t_1,t_2)$,
\begin{equation}\label{U_periodic}
U(t_1+T_{\!B},t_2+T_{\!B})=\mathcal X_{\!B}^{\,-1}\,U(t_1,t_2)\,\mathcal X_{\!B}.
\end{equation}
This suggests that the time evolution at any time can be described by the time evolution operator over the time interval $[0,T_{\!B}]$. 
Given that the reduced time $\tilde t\in[0,T_{\!B}]$ and $m\in\{0,1,2,\cdots\}$, any time $t\in[mT_{\!B},(m+1)T_{\!B}]$ can be written as $t=t_m=\tilde t+mT_{\!B}$. Therefore, 
$U(t)$ can be decomposed as
\begin{equation}\label{U_1cycle}
\begin{split}
U(t_m)&=U\big(\,\tilde t+mT_{\!B},T_{\!B}\big)\,U(T_{\!B})\\
&=\mathcal X_{\!B}^{\,-1}\,U\big(\,\tilde t+(m-1)T_{\!B})\,\mathcal X_{\!B}U(T_{\!B})\\
&=\mathcal X_{\!B}^{\,-m}\,U(\,\tilde t\,)\,\big[\mathcal X_{\!B}\,U(T_{\!B})\big]^{\,m},
\end{split}
\end{equation}
using the relation $U(t_1,t_2)=U(t_1,t_3)\,U(t_3,t_2)$ and Eq.~\eqref{U_periodic}.

Since our concern is the density matrix in the nonequilibrium steady state $\rho^{\rm NESS}_K$, let us consider the long-time limit for sufficiently large $m$ below.
In this case, the first term in Eq.~\eqref{DM} can be neglected. By decomposing the integral of the second term
as $\int_{\,t_0=0}^{\,t_m}ds=\int_{\,mT_{\!B}}^{\,\tilde t+mT_{\!B}}ds+\int_{\,0}^{\,mT_{\!B}}ds$, 
we obtain
\begin{align}\label{rhoktm}
&\rho_k(t_m)\simeq\rho_{k0}(t_m)\!+\!\rho_{k1}(t_m)=\mathcal X_{\!B}^{\,-m}\,\Big[\,\tilde\rho_{k0}(\,\tilde t\,)\!+\!\tilde\rho_{k1}(\,\tilde t\,)\,\Big]\,\mathcal X_{\!B}^{\,m}, \\
\label{MainTerm}
&\tilde\rho_{k0}(\,\tilde t\,)=U(\,\tilde t\,)\left(\int_{0}^{\,\tilde t\,}\!\!U^\dag(\,\tilde s\,)\,\Sigma(\,\tilde s\,)\,U(\,\tilde s\,)\e^{-(\,\tilde t-\,\tilde s\,\,)/\tau}d\tilde s\right)U^\dag(\,\tilde t\,),\\
\label{InterferenceTerm}
&\tilde\rho_{k1}(\,\tilde t\,)=
\!\sum_{n=0}^{m-1}\mathcal X_{\!B}^{\,n}\, U(t_n)\,\mathcal X_{\!B}\,\tilde\rho_{k0}(T_{\!B})\,\mathcal X_{\!B}^{\,-1}U^\dag(t_n)\,\mathcal X_{\!B}^{\,-n}\e^{-t_n/\tau}.
\end{align}
See Appendix~\ref{Derivation_DM} for the details. 
The terms $\tilde\rho_{k0}(\,\tilde t\,)$ and $\tilde\rho_{k1}(\,\tilde t\,)$ correspond to the components that do not include and include the interference effect of Bloch-Zener oscillations, respectively. By using Eq.~\eqref{U_1cycle}, each term in Eq.~\eqref{InterferenceTerm} can be written in the form $U(\,\tilde t\,)\,\rho U^\dag(\,\tilde t\,)\e^{-{\tilde t}/\tau}$. For example, the first term and the second term, respectively, are expressed as
\begin{subequations}\label{rho1}
\begin{align}
\tilde\rho_{k1}^{(1)}(\,\tilde t\,)&=U(\,\tilde t\,)\,\Big(\mathcal X_{\!B}\,\tilde\rho_{k0}(T_{\!B})\,\mathcal X_{\!B}^{\,-1}\Big) U^\dag(\,\tilde t\,)\e^{-{\tilde t}/\tau}, \\
\tilde\rho_{k1}^{(2)}(\,\tilde t\,)&=U(\,\tilde t\,)\,\Big(\mathcal X_{\!B}\,\tilde\rho_{k1}^{(1)}(T_{\!B})\,\mathcal X_{\!B}^{\,-1}\Big) U^\dag(\,\tilde t\,)\e^{-{\tilde t}/\tau}.
\end{align}
\end{subequations}
The physical meaning becomes clear by comparing these forms with the first term in Eq.~\eqref{DM}. The first term in Eq.~\eqref{DM}, 
$U(t)\rho_k(0)U^\dag(t)\e^{-t/\tau}$, implies that the electron in state $k$ undergoes Landau-Zener tunneling through the time evolution $U(t)$ and decays with a damping rate $\tau^{-1}$. Therefore, $\tilde\rho_{k1}^{(1)}(\,\tilde t\,)$ indicates that an electron experiencing one cycle of Bloch oscillation undergoes Landau-Zener tunneling through $U(t)$, with the resulting interference decaying at a rate of $\tau^{-1}$.
$\tilde\rho_{k1}^{(2)}(\,\tilde t\,)$ corresponds to the interference effect after one cycle of Bloch-Zener oscillation.

Note that Eq.~\eqref{InterferenceTerm} includes the decay factor $\e^{-nT_{\!B}/\tau}$.
The interference effect strongly depends on the ratio of the Bloch period $T_{\!B}$ to the relaxation time $\tau$.
Thus, in $T_B/\tau> 1$, Eq.~\eqref{rho1} gives a reliable approximation of the interference term of Eq.~\eqref{InterferenceTerm}.
In the following section, let us demonstrate that $\rho_k(t_m)$ for sufficiently large $m$ describes the distribution at the wavevector $k-E\tilde t$ in the nonequilibrium steady state,
\begin{align}\label{NESS}
\rho_k(t_m)\simeq \rho^{\rm NESS}_{k-E\tilde t}=\mathcal X_{\!B}^{\,-m}\,\tilde\rho^{\rm NESS}_{k-E\tilde t}\,\mathcal X_{\!B}^{\,m}.
\end{align}

\subsection{Nonequilibrium steady state}
As a two-band gapped system,
let us consider the Su-Schrieffer-Heeger (SSH) model, 
\begin{align}
\mathcal H_k=v\cos(ka_0/2)\sigma_x+\delta v\sin(ka_0/2)\sigma_y,\label{SSH}
\end{align}
where $\sigma_{i=x,y,z}$ are the Pauli matrices. 
This model exhibits the energy gap minima, $\Delta_k=2\delta\nu$, located at $ka_0=\pm\pi$.
Hereafter, we set $v=1,~\delta v=0.1,~\tau^{-1}=0.01$, and $Ea_0=0.02$. The temperature is set to zero.
The period of the Bloch oscillation is given by $T_B=2\pi/|E|a_0=100\pi=\pi\,\tau$.

\begin{figure}[t]
\centering
\includegraphics[width=0.9\hsize]{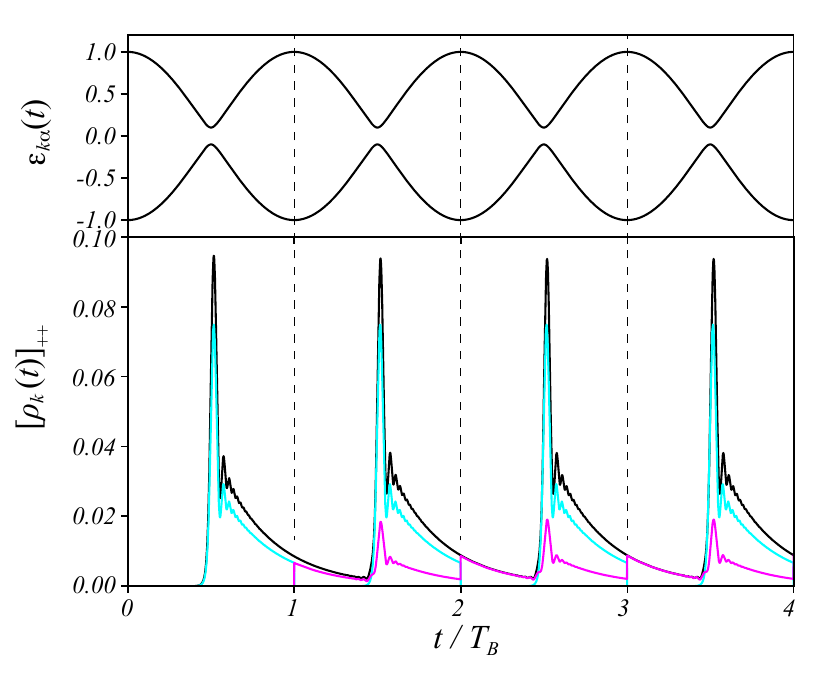}
\caption{The snapshot eigenenergy (top panel) and the upper band occupation (bottom panel) as a function of $t$ for an electron with an initial wavevector of $ka_0=2\pi$. Both oscillate periodically in $T_B$, reflecting Bloch oscillations. The upper band occupation exhibits a peak structure due to the tunneling process in the gap minima. The cyan and magenta lines correspond to $\rho_{k0}(t)$ and $\rho_{k1}(t)$, respectively. For $t>T_B$, the structure of the curve quickly converges to a periodic function with period $T_B$ due to $T_B/\tau \gg 1$, indicative of a nonequilibrium steady state. 
}\label{fig:Occ}
\end{figure}

Figure~\ref{fig:Occ} shows the time evolution of the snapshot eigenenergy (top panel) and the upper band occupation (bottom panel) for an electron with an initial wavevector of $ka_0=2\pi$. The snapshot eigenenergy exhibits periodic oscillations with a period $T_B$, reflecting Bloch oscillations. The upper band occupation exhibits a peak structure due to the tunneling process each time the electron passes through the gap minima. For $0<t<T_B$, the interference term of Eq.~\eqref{InterferenceTerm} (magenta line) is absent. The deviation between the black and cyan lines represents the contribution from the first term in Eq.~\eqref{DM}. Once $t$ exceeds $T_B$, this contribution becomes negligible.
Since subsequent tunneling processes occur before the excited electron is fully annihilated by dissipation, the excited electron never completely disappears after the first tunneling event and contributes to the interference of the subsequent tunneling events. This interference is thought to form the interference term represented by the magenta line. 
The structure of $\rho_k(t)$ quickly converges to a periodic function with period $T_B$ due to $T_B/\tau \gg 1$. 
This behavior indicates a nonequilibrium steady state. 
The difference in the initial wavevector merely shifts this structure along the $t$-axis.
Thus, the validity of Eq.~\eqref{NESS} can be confirmed.

\subsection{Current response}
In this section, let us discuss the interference of the Landau-Zener tunneling in the current response, given by
\begin{equation}\label{Current}
J(t)=-e\int_{\rm BZ}\frac{dk}{2\pi}\,\Tr\,\big[v_k(t)\rho_k(t)\big]\,.
\end{equation}
Here, $v_k(t)$ is the velocity matrix in the adiabatic basis,
\begin{equation}\label{vk}
\begin{split}
[v_k(t)]_{\alpha\beta}&=\mel{\Phi_{k\alpha}(t)}{\partial_k\mathcal H_k(t)}{\Phi_{k\beta}(t)}\\
&=
\left\{
\begin{array}{ll}
\partial_k\varepsilon_{k\alpha}(t)& (\alpha=\beta),\\[3pt]
\alpha\, i\Delta_k(t)[\mathcal W_k(t)]_{\alpha\beta}/E& (\alpha\neq\beta).
\end{array}
\right.
\end{split}
\end{equation}
The integral of Eq.~\eqref{Current} is performed over the Brillouin Zone (BZ). 
In the long-time limit, Eq.~\eqref{Current} represents the electric current in the nonequilibrium steady state,
\begin{align}
\widetilde J(E)=J(t\rightarrow \infty)=-e\int_{\rm BZ}\frac{dK}{2\pi}\Tr\big[\tilde v_K\,\tilde \rho^{\rm NESS}_K\big]\,,
\end{align}
using the relation $\tilde v_{k-E\tilde t}=v_k(\tilde t)=\mathcal X_{\!B}^{\,m}\,v_k(t_m)\,\mathcal X_{\!B}^{-m}$.

\begin{figure}[t]
\centering
\includegraphics[width=0.95\hsize]{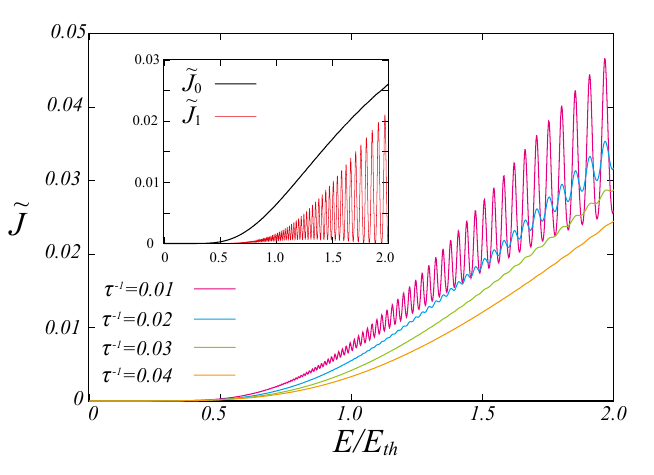}
\caption{Field dependence of the DC current for several damping $\tau^{-1}$. $E_{th}=\Delta^2/4v$ is the threshold field. The inset depicts the decomposition of contributions from $\widetilde J_0$ and $\widetilde J_1$ in $\tau^{-1}=0.01$. We can see that the oscillation at high fields in $\widetilde J$ arises from the interference term $\widetilde J_1$. Interestingly, while the amplitude of this oscillation depends on the damping $\tau^{-1}$, its period remains unaffected.} \label{fig:SSHcurrent}
\end{figure}

Figure \ref{fig:SSHcurrent} shows the field dependence of the electric current for several damping $\tau^{-1}$.
The magenta line represents the electric current with $\tau^{-1}=0.01$, which oscillates in the range of the high field. This oscillation is suppressed as the damping becomes stronger. Why does the electric current oscillate at high electric fields? 
Corresponding to the decomposition $\tilde\rho^{\rm NESS}_K=\tilde\rho_{K0}+\tilde\rho_{K1}$, $\widetilde J(E)$ can be decomposed as $\widetilde J(E)=\widetilde J_0(E)+\widetilde J_1(E)$. The inset of Fig.~\ref{fig:SSHcurrent} depicts these contributions for $\tau^{-1}=0.01$. 
One can see that the oscillation in $\widetilde J(E)$ arises from the interference term $\widetilde J_1(E)$. 
This corresponds to the Bloch-Zener oscillation in the current response. 
A similar result has been reported in Ref.~\cite{PhysRevB.85.115433}. 
Interestingly, while the amplitude of this oscillation depends on the damping $\tau^{-1}$, its period remains unaffected.
To investigate this aspect in more detail, it is essential to analyze the phase component involved in carrier generation. 
In the following section, we delve deeper into generic cases, including noncentrosymmetric systems.

\section{Geometric effects in multi-tunneling}\label{Sec:GeometricEffects}
In this section, we explore the geometric effects in multi-tunneling induced by Bloch-Zener oscillations. First, we formulate the density matrix in a nonequilibrium steady state using the adiabatic-impulse approximation (AIA)~\cite{PhysRevA.55.R2495,PhysRevA.73.063405,PhysRevResearch.4.033075}  and derive the condition for constructive interference (CCI). Finally, we elucidate the geometric effects of multi-tunneling on carrier occupation and current response.

\subsection{Overview}\label{Sec:CarrierBZO}
For simplicity, we analyze the density matrix in a nonequilibrium steady state where $T_B\gg \tau$. In this case, due to the factor $\e^{-T_B/\tau}$, the multi-tunneling effect can be adequately evaluated up to the second tunneling event. Below, we provide an overview of the formulation of the density matrix in the nonequilibrium steady state.

First, we introduce the AIA to analytically evaluate the time evolution operator $U(\tilde t)$. AIA is a reliable approximation for the time evolution operator of isolated systems and is widely used as a convenient method to describe interference arising from different tunneling processes, known as Landau-Zener-St\"uckelberg interference~\cite{IVAKHNENKO20231,SHEVCHENKO20101}. In AIA, it is assumed that only the transition at the point of minimum gap for a single electron is non-adiabatic, while all other processes remain adiabatic. 

In the adiabatic basis, the adiabatic time evolution operator, $U_{\rm ad}$, becomes an identity matrix. Therefore, the time evolution operator for the isolated system can be approximated as follows.
\begin{equation}\label{U:AIA}
U(\,\tilde t\,)=\left\{
\begin{array}{ll}
I & (\,\tilde t < t_g) \\
T & (\,\tilde t > t_g)
\end{array}
\right.,
\end{equation}
where $t_g$ denotes the time at which the tunneling process occurs, and $T$ is the transfer matrix associated with the tunneling process. See Appendix~\ref{App:AIA} for the details.
The transfer matrix $T$ can be determined based on the continuous model that describes the vicinity of the gap minima where the tunneling process occurs.
The low-energy excitations of the SSH model considered above can be described by the following Landau-Zener model,
\begin{align}\label{LZmodel}
\mathcal H_{k}=vk\sigma_x + \Delta/2\sigma_z\,.
\end{align}
Here, to account for the extension to noncentrosymmetric systems, we assume that the low-energy excitations are described by the twisted Landau-Zener model given as
\begin{align}\label{TLZmodel}
\mathcal H_k=vk\sigma_x+\eta k^2/2\sigma_y+\Delta/2\sigma_z\,.
\end{align}
In this case, the transfer matrix in the adiabatic basis is given by
\begin{equation}\label{TransferMatrix_LZ}
   T=	\begin{pmatrix}
	\sqrt{1-P\,}\e^{-i\phi_{S}}  &  -\sqrt{P\,} \e^{i\delta\theta_{k}(t_g)}\\
	\sqrt{P\,}\e^{-i\delta\theta_{k}(t_g)}  &  \sqrt{1-P\,} \e^{i\phi_{S}}
	\end{pmatrix},
\end{equation} 
where $\delta\theta_{k}(t)=\theta_{k+}(t)-\theta_{k-}(t)$ and $P$ denotes the tunneling probability $P=\e^{-2\pi\delta}$ with
\begin{align}
\delta=\frac{\Big(\Delta+\frac{E\eta}{2v}\Big)^2}{8v|E|}.
\end{align}
The phase $\phi_S$, known as the Stokes phase~\cite{PhysRevA.55.R2495}, represents the phase acquired during the tunneling process and is expressed as
\begin{align}
\phi_{S}=\frac{\pi}{4}+\arg\,\Gamma(1-i\delta)+\delta\,(\ln\delta-1),
\end{align}
where $\Gamma(x)$ is the gamma function.
See Appendix~\ref{App:AIA_TLZ} for the details.
For finite values of $\eta$, $\delta$ exhibits nonreciprocity and demonstrates perfect tunneling at a specific electric field $E$ where $\delta = 0$. Furthermore, when $E$ is sufficiently large, $\delta$ increases proportionally with $E$, indicative of the counter-diabatic driving.

\subsection{Condition for constructive interference (CCI)}
Let us proceed to the evaluation of $\tilde \rho_k(\,\tilde t\,)$. For $T_B\gg\tau$, it is approximately described by Eq.~\eqref{rho1}. For simplicity, we evaluate them within the RTA framework to derive the interference conditions, as the DPA corrections are primarily intended to improve the gapped behavior in the low-field regime and are not essential to the present objective. In this case, the diagonal parts

$[\tilde \rho_k(\,\tilde t\,)]_{\alpha\alpha}$ is estimated as
\begin{align}\label{NESS_AIA}
[\tilde\rho_k(\,\tilde t\,)]_{\alpha\alpha}\simeq f_D(\varepsilon_{k\alpha}(\,\tilde t\,))+\alpha\, N(E)\,g(\,\tilde t\,)\e^{-(\tilde t-t_g)/\tau},
\end{align}
where $g(\,\tilde t\,)=\Theta(\tilde t-t_g)+\e^{-T_B/\tau}\Theta(t_g-\tilde t)$ and the carrier generation $N(E)=N_1(E)+N_2(E)$ with
\begin{subequations}
\begin{align}
&N_1(E)\simeq P\, \delta f_k(t_g),\\
&N_2(E)\simeq \bigg[-P+4P(1-P)\cos^2\left(\frac{\Theta_B(E)+2\phi_S}{2}\right)\bigg]\,\delta f_k(t_g)\e^{-T_B/\tau}.
\end{align}
\end{subequations}
\noindent
Here, $\Theta$ is the Heaviside step function. 
Given that $\e^{-T_B/\tau}$ is negligible for $T_B\gg\tau$, we obtain
\begin{align}
[\tilde \rho_k(\,\tilde t\,)]_{\alpha\alpha}\simeq f_D(\varepsilon_{k\alpha}(\,\tilde t\,))+N_1(E)\e^{-(t'-t_g)/\tau}\Theta(t'-t_g),
\end{align}
which is consistent with the result of the Green's function approach~\cite{PhysRevB.102.245141} for a continuous model with a single tunneling event. 
Interference effects are incorporated into the multi-tunneling effect in carrier generation, such as $N_2(E)$.
The upper band occupation is enhanced under a specific electric field that satisfies the condition for constructive interference (CCI), given by
\begin{align}
\label{condition1}
&(\Delta_{\rm{av}}+ER_{\rm{av}})\frac{2\pi}{|E|a_0}+2\phi_S(E)=2n\pi~~(n\in\mathbb{Z}),\\
&\Delta_{\rm av}=\frac{1}{T_B}\int_{0}^{T_B}\Delta_k(s)ds=a_0\int_{\rm BZ}\frac{dk}{2\pi}\Delta_{k},\\
&R_{\rm av}=\frac{1}{T_B}\int_{0}^{T_B}R_k(s)ds=a_0\int_{\rm BZ}\frac{dk}{2\pi}R_{k},
\end{align}
where $\Delta_{\rm av}$ and $R_{\rm{av}}$ are the momentum average of the energy gap and the shift vector, respectively. 
Note that the CCI is independent of the damping $\tau^{-1}$. 
This explains why the oscillation period in Fig.~\ref{fig:SSHcurrent} is independent of the damping.
While $R_{\rm av}=0$ in this case,
in the following, we investigate the nonreciprocity in tunneling probability and electric current in noncentrosymmetric systems with finite $R_{\rm av}$.

\subsection{Nonreciprocal transport}\label{Sec:Nonreciprocity}
As a simple model for noncentrosymmetric systems, we consider a Rice-Mele model,
\begin{align}\label{RiceMele}
\mathcal{H}_k=v\cos(ka_0/2)\sigma_x+\delta v\sin(ka_0/2)\sigma_y+m\sigma_z\,.
\end{align}
This model includes the SSH model in Eq.~\eqref{SSH} as a special case when $m=0$, and its minimal band gap is given by $\Delta=2\sqrt{\delta v^2+m^2}$ (for $\delta v<v$). 
In the analysis of the CCI, the low-energy excitations of this model are mapped onto the twisted Landau-Zener model in Eq.~\eqref{Ham_TLZ}, with $\eta=\delta v\,m/\Delta$.
Hereafter, we set $v=1,~\delta v=0.1,~m=0.2$, and $\tau^{-1}=0.02$.

\begin{figure}[t]
\centering
\includegraphics[width=0.9\hsize]{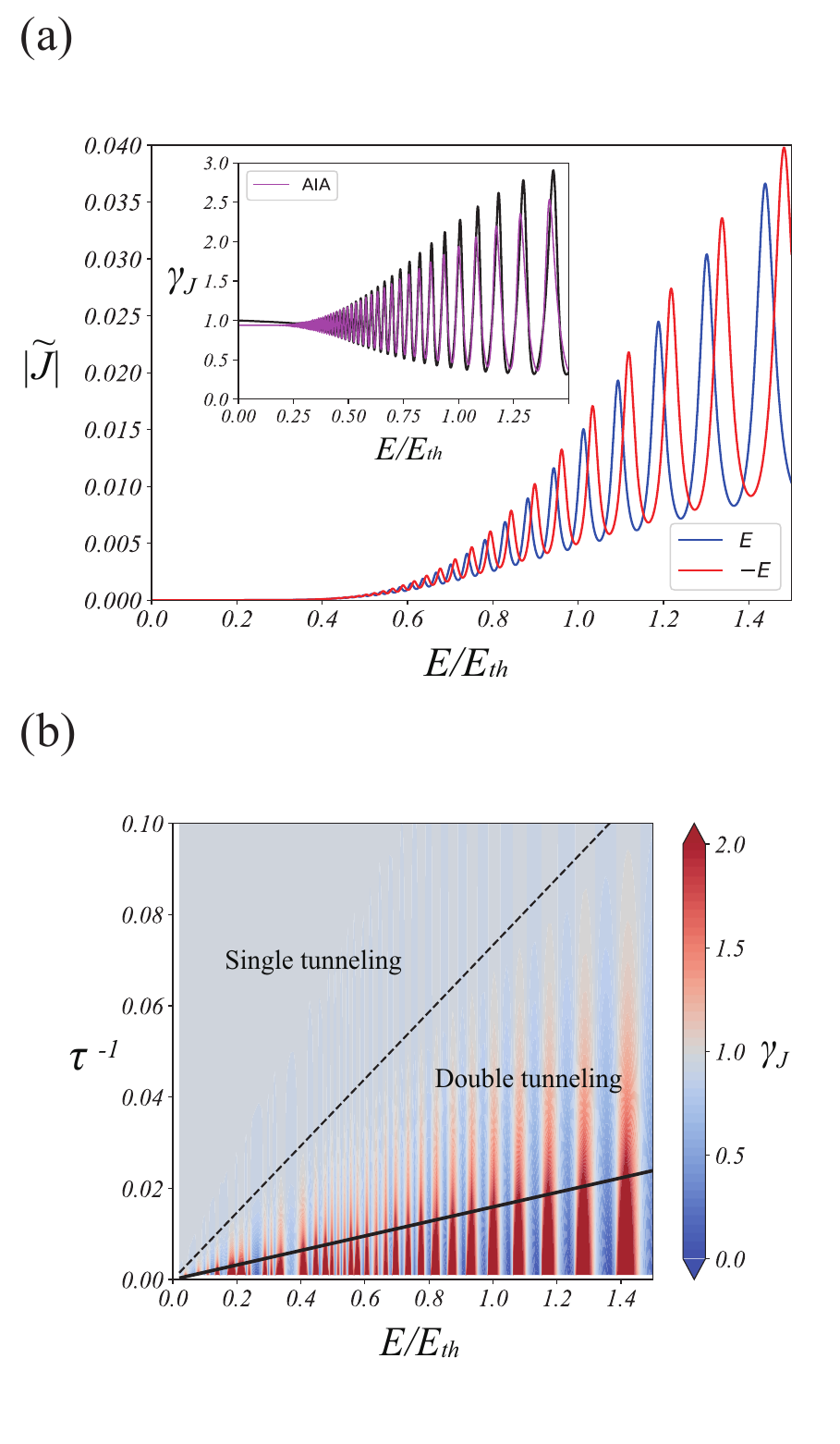}
\caption{The nonreciprocal transport in the Rice-Mele model. (a) (Main panel) Field-dependence of the electric current. $E_{th}=\Delta^2/4v$ is the threshold field. (Inset) Field-dependence of the nonreciprocity ratio of the electric current. The black and magenta lines correspond to the original result and the approximation by Eq.~\eqref{ratio_J}, respectively. (b) Contour plot of Eq.~\eqref{ratio_J} as a function of electric field and damping. The dash and solid lines, respectively, correspond to $\e^{-T_B/\tau}=0.01$ and $\e^{-T_B/\tau}=\e^{-1} $\big(\,$E\tau= 2\pi/a_0$\,\big).}\label{fig:RMcurrent}
\end{figure}

The main panel of Fig.~\ref{fig:RMcurrent}(a) shows the field dependence of the electric current $\widetilde J$ in Eq.~\eqref{Current}.
The blue and red lines correspond, respectively, to the electric current for $E>0$ and $E<0$, indicating the nonreciprocal current.
In the high-field regime, the peak positions of the current depend on the direction of the applied electric field.
This directional dependence arises from the presence of the shift vector.
The inset of Fig.~\ref{fig:RMcurrent}(a) represents the nonreciprocity ratio of the electric current, $\gamma_{J}(E)=|\widetilde J(E)/\widetilde J(-E)|$.
Notably, the nonreciprocity ratio oscillates above unity as the field intensity increases, attributed to the interference effect. This behavior implies that the directionality of the current can be controlled by tuning the field intensity.

Let us proceed with an analysis similar to that of the CCI in the previous section to investigate the effects of multi-tunneling and damping on the nonreciprocity ratio. For simplicity, we neglect the less essential interband current and consider only the contribution from the intraband current~\cite{PhysRevB.102.245141}. 
In this case, $\widetilde J(E)$ can be evaluated as
\begin{align}
\widetilde J(E) &=-\int_{\rm BZ}\frac{dK}{2\pi}\sum_{\alpha=+,-}\alpha\frac{\partial \varepsilon_{K\alpha}}{\partial K}[\rho(K)]_{\alpha\alpha}
=-\int_{\rm BZ}\frac{\partial \Delta_K}{\partial K}[\rho(K)]_{++} \nonumber \\
&\simeq \mathrm{sgn}(E)\frac{(E\tau)^2}{2\pi}\left.\frac{\partial^2 \Delta_K}{\partial K^2}\right|_{K=\frac{\pi}{a_0}}N(E)\,(1-\e^{-T_B/\tau}),
\end{align}
and the nonreciprocity ratio is approximately given by 
\begin{equation}\label{ratio_J}
\gamma_J(E)\simeq \frac{N(E)}{N(-E)}=\frac{N_1(E)+N_2(E)}{N_1(-E)+N_2(-E)}.
\end{equation}
This returns to the usual ratio of the tunneling probabilities, $\gamma_{P}=P(E)/P(-E)$, if the interference term containing $\e^{-T_B/\tau}$ can be neglected under the condition $T_B\gg\tau$. 
In the inset of Fig.~\ref{fig:RMcurrent}(a), one can verify that Eq.~\eqref{ratio_J} (magenta) well reproduces the original result (black).

In Fig.~\ref{fig:RMcurrent}(b), we shows the contour map of Eq.~\eqref{ratio_J} as a function of electric field and damping. 
For $T_B\gg \tau$, i.e., $|E|\tau \ll 2\pi/a_0$, we find $\gamma_J\simeq \gamma_P$, and the oscillating behavior caused by the interference does not appear. Next, for $T_B>\tau$, we observe pronounced oscillations and significant nonreciprocity of the electric current, attributed to the interference effect caused by the second tunneling event.
\begin{figure}[t]
\centering
\includegraphics[width=0.9\hsize]{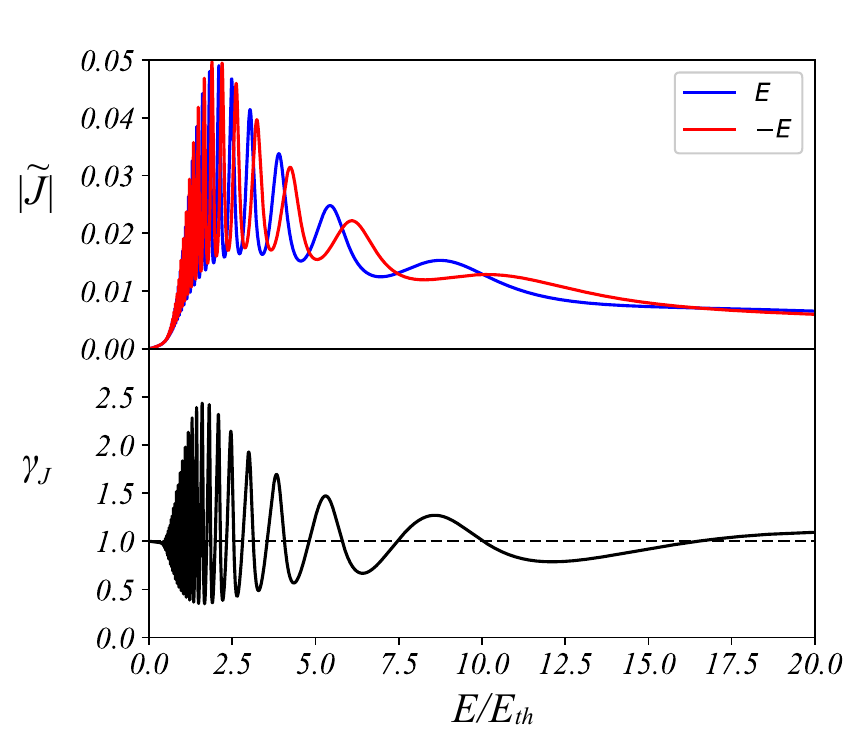}
\caption{(a) The nonreciprocal current in the RTA under much higher fields and (b) the correspoinding nonreciprocity ratio. The oscillation and magnitude of electric current is suppressed with increasing field intensity. As a result, the nonreciprocity ratio asymptotically approaches unity.}\label{fig:Highfield}
\end{figure}

Finally, let us briefly consider the case of $T_B<\tau$, i.e., $|E|\tau > 2\pi/a_0$. Figure \ref{fig:Highfield}  depicts the RTA electric current and its nonreciprocity at $\tau^{-1}=0.02$ under a higher field. 
Contrary to the above results, in a higher electric field, both the magnitude and oscillations of the current are suppressed with increasing field intensity, and the nonreciprocity ratio asymptotically approaches unity. This behavior is likely caused by the cancellation 
of oscillations due to multi-tunneling effects and Joule heating~\cite{PhysRevB.87.085119}, or dynamic localization in the high-field regime. 

\section{Discussion and Conclusion}\label{Sec:Conclusion}
In this paper, we have investigated the role of quantum geometric effects and the multi-tunneling effects in nonreciprocal Landau-Zener tunneling. We first calculated the density matrix of lattice systems based on the quantum kinetic equation, decomposing the dissipative dynamics into coherent and incoherent components to clarify their contributions. The derived density matrix characterizes the nonequilibrium steady state of insulating systems, facilitating efficient calculation of nonequilibrium steady state physical quantities with low computational cost. In noncentrosymmetric systems, we demonstrated that quantum geometric effects enhance the nonreciprocity of current induced by Bloch-Zener oscillations, allowing for controlled directionality of the current.

Bloch oscillations have been experimentally observed in optical lattice systems, suggesting that our results may also be observable. 
Although observing Bloch oscillations in lattice systems is generally challenging, they have been successfully observed in semiconductor superlattices~\cite{PhysRevB.46.7252} and proposed in moiré materials~\cite{PhysRevLett.126.256803}. In particular, moir\'e materials form a flat band near the “magic angle”, enhancing quantum geometric effects~\cite{doi:10.1126/science.aaw3780,doi:10.1126/science.aay5533,PhysRevResearch.4.013164}.
Thus, systems such as twisted bilayer graphene offer a promising platform for exploring intriguing phenomena driven by geometric multi-tunneling. 

Experimentally, generating strong electric fields may be more practical with AC fields than with DC fields.
Our findings may be potentially observed in gapped 2D Dirac systems under a circular polarized light, which can describe approximately the Rice-Mele model under the DC electric field discussed in this paper, by replacing $ka_0/2$ to $\Omega t$ and $v,\delta v$ to $vE_0/\Omega$ in Eq.~\eqref{RiceMele}.

Furthermore, it would be both significant and worthwhile to extend this study to AC electric fields. Under AC fields, the dominant process—whether multiphoton absorption or quantum tunneling—continuously changes via the Keldysh crossover. 
In this context, nonreciprocal Landau-Zener tunneling and shift currents represent extreme cases of phenomena arising from the same geometric quantity, i.e. quantum tunneling and multiphoton absorption. Elucidating these relationships in terms of the Keldysh crossover could lead to a systematic understanding of quantum geometric effects in the nonperturbative regime.

\begin{acknowledgments}
We are grateful to A. Daido for valuable comments. This work was supported by KAKENHI Grants No. 19H01842, No. 23K25827, No. 20K14407, No. 23K25816, No. 24K01333.
\end{acknowledgments}

\appendix
\section{Density matrix in nonequilibrium steady state}\label{Derivation_DM}
Let us consider the long-time limit of Eq.~\eqref{DM}.
In this limit, the first term vanishes due to the decay factor $\e^{-t/\tau}$, and Eq.~\eqref{DM} becomes
\begin{equation}\label{DM_longtime}
\rho_k(t)=U(t)\left(\int_{0}^t U^\dag(s)\Sigma(s)U(s)\e^{-(t-s)/\tau}~ds\right)U^\dag(t).
\end{equation}
This is divided into the following two parts,
\begin{align}\label{DM1}
&\rho_{k0}(t_m)=U(t_m)\left(\int_{mT_B}^{t_m} U^\dag(s)\Sigma(s)U(s)\e^{-(t_m-s)/\tau}~ds\right)U^\dag(t_m),\\
\label{DM2}
&\rho_{k1}(t_m)=U(t_m)\left(\int_{0}^{mT_B} U^\dag(s)\Sigma(s)U(s)\e^{-(t_m-s)/\tau}~ds\right)U^\dag(t_m),
\end{align}
where $t_m=\,\tilde t\,+mT_B$ with the reduced time $\tilde t\in[0,T_B]$.

Using $U(t_m)U^\dagger(s_m)=\mathcal{X}_B^{-m}U(\tilde{t})U^\dagger(\tilde{s})\mathcal{X}_B^m$ [See Eq.~\eqref{U_1cycle}] and $\Sigma(t_m)=\mathcal X^{-m}_B\Sigma(\,\tilde t\,)\mathcal X^{m}_B$, Eq.~\eqref{DM1} becomes
\begin{align}
&\rho_{k0}(t_m)=U(t_m)\left(\int^{\tilde t}_{0}U^\dag(s_m)\Sigma(s_m)U(s_m)\e^{-(t_m-s_m)/\tau}d\tilde s\right)U^\dag(t_m) \nonumber \\
&=\mathcal X_B^{-m}U(\,\tilde t\,)\left(\int_{0}^{\,\tilde t\,}U^\dag(\tilde s)\Sigma(\tilde s)U(\tilde s)\e^{-(\,\tilde t\,-\tilde s)/\tau}d\tilde s\right)U^\dag(\,\tilde t\,)\mathcal X_B^{m} \nonumber \\
&=\mathcal X_B^{-m}\tilde \rho_{k0}(\,\tilde t\,)\mathcal X_B^{m}.
\end{align}
On the other hand, using
\begin{align}
U(t_m)U^\dagger(s_{m-n})&=\mathcal{X}_B^{-m+n}U(t_n)U^\dagger(\tilde s)\mathcal{X}_B^{m-n}\nonumber\\&=\mathcal{X}_B^{-m+n}[\mathcal{X}_B^{-1}U(t_{n-1})\mathcal{X}_BU(T_B)]U^\dagger(\tilde s)\mathcal{X}_B^{m-n},
\end{align}
Eq.~(\ref{DM2}) is calculated as
\begin{widetext}
\begin{equation}
\begin{split}
\rho_{k1}(t_m)&=U(t_m)\sum_{n=1}^{m}\left(\int_{(m-n) T_B}^{(m-n+1)T_B}U^\dag(s)\Sigma(s)U(s)\e^{-(t_m-s)/\tau}ds\right)U^\dag(t_m)\\
&=U(t_m)\sum_{n=1}^{m}\left(\int_{0}^{T_B}U^\dag(s_{m-n})\Sigma(s_{m-n})U(s_{m-n})\e^{-(t_n-\tilde s)/\tau}d\tilde s\right)U^\dag(t_m)\\
&=\sum_{n=1}^{m}\mathcal{X}_B^{-m+n-1}U(t_{n-1})\mathcal{X}_BU(T_B)\left(\int_{0}^{T_B}U^\dagger(\tilde s)\Sigma(\tilde s)U(\tilde s) \e^{-(t_n-\tilde s)/\tau}d\tilde s\right)U^\dagger(T_B)\mathcal{X}_B^{-1}U^\dagger(t_{n-1})\mathcal{X}_B^{m-n+1}\\
&=\mathcal X_B^{-m}\left(\sum_{n=0}^{m-1}\mathcal X_B^{n}U(t_n)\mathcal X_B\rho_{k,0}(T_B)\mathcal X^{-1}_BU^\dag(t_n)\mathcal X_B^{-n}\e^{-t_n/\tau}\right)\mathcal X_B^m
=\mathcal X_B^{-m}\tilde\rho_{k1}(\,\tilde t\,)\mathcal X_B^m.
\end{split}
\end{equation}
\end{widetext}
$\rho_{k1}(\,\tilde t\,)$ represents the multi-tunneling process because $U(t_l)$ and $\rho_{k0}(T_B)$ contain the knowledge about different tunneling processes.
Finally, we obtain Eqs.~\eqref{rhoktm}, \eqref{MainTerm} and \eqref{InterferenceTerm}.

\section{Adiabatic impulse approximation in adiabatic basis}\label{App:AIA}
We introduce the adiabatic impulse approximation (AIA) in Landau-Zener problem.
In the AIA, we consider the Houston basis, $\ket{\Psi(t)}=\sum_{\alpha}c_{\alpha}(t)\ket{u_{\alpha}(t)}$ with initial state $\ket{u_{-}(t_0)}$.
This basis differs from the adiabatic basis only in its dynamical and Berry phases.
The time evolution of $\bm{c}(t)=(c_{+}(t),c_{-}(t))^T$ is governed by the time evolution operator in the Houston basis,
\begin{align}
&U^H(t,t_0)=\mathcal{T}\exp\bigg[-i\int_{t_0}^t\mathcal{H}^H(\,\tilde t\,)d\,\tilde t\,\bigg],\\
&\mathcal{H}^{H}(t)=V^{H\dag}(t)\bigg[\mathcal{H}(t)-i\frac{\partial}{\partial t}\bigg]V^{H}(t),
\end{align}
where $V^{H}(t)=(\ket{u_{+}(t)},\ket{u_{-}(t)})$.
In the AIA, the time evolution is devided
into three cases: (i) $t<t_g$, (ii) $t=t_g$, and (iii) $t>t_g$, where $t_g$ is the time at which the electron passes through the gap minima.
First, in the case (i), the time evolution of $\bm{c}(t)$ is expressed as $\bm{c}(t)=U_{ad}(t,t_0)\bm{c}(t_0)$, where $U_{ad}(t_1,t_2)$ is the adiabatic time evolution 
\begin{align}
&U_{ad}(t_1,t_2)=
\left(
\begin{matrix}
\e^{-i\theta_{k+}(t_1,t_2)}&0\\
0&\e^{-i\theta_{k-}(t_1,t_2)}
\end{matrix}
\right),\\
&\theta_{k\alpha}(t_1,t_2)=
\int_{t_2}^{t_1}[\varepsilon_{k\alpha}(\,\tilde t\,)+EA_{k,\alpha\alpha}(\,\tilde t\,)]d\,\tilde t\,.
\end{align}
Next, at (ii) $t=t_g$, a Landau-Zener transition is assumed to occur, imposing the relation, $\bm{c}(t_g+)=T^H\bm{c}(t_g-)$, 
where $T^H$ is the transfer matrix in the Houston basis. 
Finally, in the case (iii), $\bm{c}(t)$ undergoes adiabatic time evolution once more, resulting in $\bm{c}(t)=U_{ad}(t,t_g)T^HU_{ad}(t_g,t_0)\bm{c}(t_0)$. 
As a result, the time evolution operator can be approximated as
\begin{equation}\label{App:TimeOp_AIA_Houston}
U^{H}(t,t_0)=\left\{
\begin{array}{ll}
U_{ad}(t,t_0) & (t < t_g) \\
U_{ad}(t,t_g+)T^HU_{ad}(t_g-,t_0) & (t > t_g)
\end{array}
\right.\,.
\end{equation}
Using $U_{ad}(t,t_0)$, a unitary transformation can convert the Houston basis onto the adiabatic basis. The Hamiltonian in the adiabatic basis and the time evolution are defined by
\begin{align}
&\mathcal{W}_k(t)=U^{\dag}_{ad}(t,t_0)\,\bigg(\mathcal{H}^{H}(t)-i\frac{\partial}{\partial t}\bigg)\,U_{ad}(t,t_0),\\
&U(t,t_0)=U^{\dag}_{ad}(t,t_0)\,U^H(t,t_0).
\end{align}
In this case, the time evolution operator in the adiabatic basis is expressed as
\begin{equation}\label{App:TimeOp_AIA}
U(t,t_0)=\left\{
\begin{array}{ll}
I & (t < t_g) \\
T =U^{\dagger}_{ad}(t_g,t_0)T^HU_{ad}(t_g,t_0) & (t > t_g)
\end{array}
\right..
\end{equation}

\section{AIA in the twisted Landau-Zener model}\label{App:AIA_TLZ}
The twisted Landau-Zener (TLZ) model,
\begin{align}\label{Ham_TLZ}
\mathcal{H}_{TLZ}(t)&=vk\sigma_x+\frac{\eta}{2}k^2\sigma_y+\frac{\Delta}{2}\sigma_z,
\end{align}
can be mapped onto the Landau-Zener model through the unitary transformation $V_{\varphi}(t)=\exp[-i\varphi(t)/2\cdot\sigma_z]$ with $\varphi(t)=\tan^{-1}[\eta k/2v]$~\cite{10.21468/SciPostPhys.11.4.075}, as follows.
\begin{align}
\mathcal{H}_{LZ}(t)
&=V^{\dag}_{\varphi}(t)\,\bigg[\mathcal{H}_{TLZ}(t)-i\frac{\partial}{\partial t}\bigg]\,V_{\varphi}(t) \nonumber \\
&=\sqrt{(vk)^2+\left(\frac{\eta k^2}{2}\right)^2}\sigma_{x}+\frac{1}{2}\left(\Delta+\frac{E\eta/2v}{1+\left(\frac{\eta k}{2v}\right)^2}\right)\,\sigma_z \nonumber \\
\label{TLZtoLZ}
&\simeq vk\sigma_x+\frac{1}{2}\bigg(\Delta+\frac{E\eta}{2v}\bigg)\,\sigma_z 
\end{align}
%
In the Houston basis, the LZ model and TLZ model are related by the following relation,
\begin{align}
&\mathcal{H}^{H}_{TLZ}(t)=V^{H\dag}(t)\left[\mathcal{H}^{H}_{LZ}(t)-i\frac{\partial}{\partial t}\right]V^H(t),\\
&V^H(t)=V^{H\dag}_{LZ}(t)V^{\dag}_{\varphi}(t)V^{H}_{TLZ}(t)
\end{align}
where $V^{H}_{LZ/TLZ}(t)$ are the  unitary matrices that diagonalize each Hamiltonian $\mathcal H_{LZ/TLZ}(t)$.
Thus, the time evolution operator of the TLZ model is expressed as
\begin{equation}
U^{H}_{TLZ}(t,t_0)=V^{H\dag}(t)U^{H}_{LZ}(t,t_0)V^{H}(t_0).
\end{equation}
In the AIA, $U^{H}_{LZ}(t,t_0)$ in the right-hand side is given by Eq.~(\ref{App:TimeOp_AIA_Houston}).
With a suitable gauge choice for $V_{LZ}^H(t)$ where $\theta_{k\alpha}(t_1,t_2)$ coincides for $\mathcal{H}_{LZ}$ and $\mathcal{H}_{TLZ}$, we can compute the unitary matrix $V^H(t)$ as
\begin{equation}
V^{H}(t)=I+\dfrac{vE\sin\varphi}{\Delta_{k}^{2}}\begin{pmatrix}0 & e^{-i\varphi}\\
-e^{i\varphi} & 0
\end{pmatrix}+O(E^{2})
\end{equation}
with $\Delta_{k}=2\sqrt{(vk)^{2}+(\eta k^{2}/2)^{2}+(\Delta/2)^{2}}$. Hereafter we neglect the second term as a subleading contribution for simplicity and take $U^{H}_{TLZ}(t,t_0)\simeq U^{H}_{LZ}(t,t_0)$.

Proceeding similarly to the formulation in Appendix~\ref{App:AIA}, the time evolution operator of the TLZ model in the adiabatic basis is expressed as 
\begin{equation}
U^{S}_{TLZ}(t)\simeq 
\left\{
\begin{array}{ll}
I & (t < 0) \\
T_{TLZ}=U^{\dag}_{ad}(0,t_0)T^H_{LZ}U_{ad}(0,t_0) & (t > 0)
\end{array}
\right.
\end{equation}
with $t_0\rightarrow-\infty$.
The transfer matrix in the adiabatic basis is expressed as
\begin{equation}
T_{TLZ}=\left(
\begin{matrix}
\sqrt{1-P_{\rm{TLZ}}}\e^{-i\phi_{S}}&-\sqrt{P_{\rm{TLZ}}}\e^{i\delta\theta_k^{TLZ}(0)}\\
\sqrt{P_{\rm{TLZ}}}\e^{-i\delta\theta_k^{TLZ}(0)}&\sqrt{1-P_{\rm{TLZ}}}\e^{i\phi_{S}}
\end{matrix}
\right),
\end{equation}
where $P_{TLZ}=\e^{-2\pi\delta},~\delta=(\Delta+E\eta/2v)^2/8v|E|$ and $\delta\theta_k^{TLZ}(t)=\theta^{TLZ}_{k+}(t,t_0)-\theta^{TLZ}_{k-}(t,t_0)$.

\section{Map a lattice model onto a continuous model}
Following Ref.~\cite{CommunPhys.1.63}, the low-energy excitations in the Rice-Mele model of Eq.~\eqref{RiceMele} can be mapped onto the following continuous model,
\begin{equation}
\begin{split}
\mathcal H_k&=-vk\sigma_x+\delta v\sigma_y+m\sqrt{1+k^2}\sigma_z\\
&\sim -vk\sigma_x+\delta v\sigma_y+m(1+k^2/2)\sigma_z.
\end{split}
\end{equation}
Using $V_{\gamma}=\e^{i(\pi/2-\gamma)/2\cdot\sigma_x}$ and $\gamma=\tan^{-1}(m/\delta v)$, 
this Hamiltonian is transformed into the twisted Landau-Zener model of Eq.~\eqref{TLZmodel}, as follows.
\begin{equation}\label{RM_continuous}
\begin{split}
\mathcal{H}_k\rightarrow&\mathcal{V}^{\dag}_{\gamma}\mathcal{H}_k\mathcal{V}_{\gamma}\\
&\simeq-vk\sigma_x-\frac{\eta}{2}k^2\sigma_y+\frac{\Delta}{2}\sigma_z,
\end{split}
\end{equation}
where $\Delta=2\sqrt{\delta v^2+m^2}$ and $\eta=2\delta vm/\Delta$. Equation~\eqref{RM_continuous} reproduces the TLZ Hamiltonian in Eq.~\eqref{TLZmodel} by replacing $-v\rightarrow v$ and $-\eta\rightarrow \eta$. For $m=0$, the Rice-Mele model is reduced to the SSH model. The corresponding continuous model becomes the LZ model.

\section{Density matrix in AIA}
Under AIA and RTA, we calculate the density matrix, $\rho_k(t_m)=\mathcal{X}_B^{-m}\big[\tilde\rho_{k0}(\,\tilde t\,)+\tilde\rho_{k1}(\,\tilde t\,)\big]\mathcal{X}_B^{m}$ 
with $\tau \ll T_B$.
Since the time evolution operator in AIA is approximated as 
\begin{equation}\label{APP:U_AIA}
\big[U(\,\tilde t\,)\big]_{\alpha\beta}=\delta_{\alpha\beta}\Theta(t_g-\,\tilde t\,)+\big[T\big]_{\alpha\beta}\Theta(\,\tilde t\,-t_g),
\end{equation}
we obtain 
\begin{widetext}
\begin{align}
&[\tilde\rho_{k0}(\,\tilde t\,)]_{\pm\pm}=F(\varepsilon_{k\pm}(\,\tilde t\,))\pm P\,\delta F_k(t_g)\e^{-(\,\tilde t\,-t_g)/\tau}\Theta(\,\tilde t\,-t_g)
-\bigg\{F(\varepsilon_{k\pm}(0))\pm P\,\delta F_k(0)\,\Theta(\,\tilde t\,-t_g)\bigg\}\e^{-\,\tilde t\,/\tau},
\\
&[\tilde\rho_{k0}(\,\tilde t\,)]_{+-}=-\sqrt{P(1-P)}\,\e^{i\delta\theta_k(t_g)-i\phi_S}\delta F_k(t_g)\e^{-(\,\tilde t\,-t_g)/\tau}\Theta(\,\tilde t\,-t_g)
+\sqrt{P(1-P)}\,\e^{i\delta\theta_k(t_g)-i\phi_S}\delta F_k(0)\e^{-\,\tilde t\,/\tau}\Theta(\,\tilde t\,-t_g),
\\
\label{App:n=0}
&\tilde\rho^{(0)}_{k1}(\,\tilde t\,)=\mathcal{X}_B\,\rho_{k0}\,(T_B)\,\mathcal{X}^{-1}_B\e^{-\,\tilde t\,/\tau}\Theta(t_g-\,\tilde t\,)
+T\mathcal{X}_B\,\tilde\rho_{k0}(T_B)\,\mathcal{X}^{-1}_B\,T^\dag\e^{-\,\tilde t\,/\tau}\Theta(\,\tilde t\,-t_g),
\\
\label{App:n=1}
&\tilde\rho^{(1)}_{k1}(\,\tilde t\,)=\mathcal{X}_B\,T\,\mathcal{X}_B\,\rho_{k0}(T_B)\,\mathcal{X}^{-1}_B\,T^\dag\mathcal{X}^{-1}_B\e^{-(\,\tilde t\,+T_B)/\tau}\Theta(t_g-\,\tilde t\,)
+(T\mathcal{X}_B)^2\rho_{k0}(T_B)\,(\mathcal{X}^{-1}_BT^\dag)^2\e^{-(\,\tilde t\,+T_B)/\tau}\Theta(\,\tilde t\,-t_g),
\end{align}
\end{widetext}
where $F(\varepsilon_{k\alpha})$ is the distribution function based on the Boltzman transport theory,
\begin{equation}
F(\varepsilon_{k\alpha}(t))=f_D(\varepsilon_{k\alpha}(t))+E\tau\frac{\partial f_D(\varepsilon_{k\alpha}(t))}{\partial k}+\cdots,
\end{equation}
$\delta F_k=F(\varepsilon_{k-})-F(\varepsilon_{k+})$ and $P$ is the tunneling probability. 
For $\,\tilde t\, > t_g$, $\tilde\rho_{k0}(\,\tilde t\,)$ includes the information about a single tunneling process due to the nonperturbative corrections, namely the terms involving $P$.

The second term in Eq.~\eqref{App:n=0} and the first term in Eq.~\eqref{App:n=1} describe a single tunneling process by the pair of $T$ and $T^\dag$, which is different from that in $\tilde \rho_{k0}(T_B)$. 
Therefore, these terms represent a double tunneling event, namely the interference between different Landau-Zener tunneling events.
Similarly, the second term in Eq.~\eqref{App:n=1} represents a triple tunneling event. 
This term is negligible due to $\tau \ll T_B$.

From the above, the distribution function in the nonequilibrium steady state, $[\rho^{\rm NESS}_{k}(\,\tilde t\,)]_{\pm\pm}$, is estimated as
\widetext
\begin{equation}\label{App:Dia_AIA}
\begin{split}
[\rho^{\rm NESS}_{k}(\,\tilde t\,)]_{\pm\pm}
&\simeq[\rho_{k,0}(\,\tilde t\,)]_{\pm\pm}+[T\mathcal{X}_B\rho_{k,0}(T_B)\mathcal{X}^{-1}_BT^\dag]_{\pm\pm} \e^{-\,\tilde t\,/\tau}\Theta(\,\tilde t\,-t_g)\\
&~~~~~~~~~~~~~~
+\bigg\{[\mathcal{X}_B\rho_{k,0}(T_B)\mathcal{X}^{-1}_B]_{\pm\pm}\e^{-\,\tilde t\,/\tau}
+[\mathcal{X}_BT\mathcal{X}_B\rho_{k,0}(T_B)\mathcal{X}^{-1}_BT^\dag\mathcal{X}^{-1}_B]_{\pm\pm}\e^{-(\,\tilde t\,+T_B)/\tau}\bigg\}\Theta(t_g-\,\tilde t\,),\\
&= F(\varepsilon_{k\pm}(\,\tilde t\,))\pm P\bigg\{1+\bigg(4\cos^2(\Theta_B/2+\phi_S)-1\bigg)\e^{-T_B/\tau}\bigg\}\delta F_k(t_g)\{\e^{-(\,\tilde t\,-t_g)/\tau}\Theta(\,\tilde t\,-t_g)+\e^{-(\,\tilde t\,+T_B-t_g)/\tau}\Theta(t_g-\,\tilde t\,)\}\\
&~~~~~~~~~~~~~~~~~~~~~~~~~~~~~~~~~~~~~~~~~~~~~~~~~~~~~~~~~~~~~~~~~~~~~~~~~~~~~~~~~~~~~~~~~~
+\mathcal{O}(P^2)+\bigg(\mbox{the terms involving $\delta F_k(t_0=0)$}\bigg).
\end{split}
\end{equation}
\endwidetext
The density matrix in the nonequilibrium steady state is independent of the initial time $t_0$, except for the time at which a sharp peak structure appears. Therefore, some terms involving $\delta F_k(t_0)$ in Eq.~\eqref{App:Dia_AIA} are expected to be canceled out by the second term in Eq.~\eqref{App:n=1} or by the terms for $n \geq 2$ in Eq.~\eqref{InterferenceTerm}. On the other hand, the off-diagonal part of $\rho^{\rm NESS}_k(\,\tilde t\,)$ is estimated as
\widetext
\begin{equation} 
\begin{split}
\big[\rho^{\rm NESS}_{k}(\,\tilde t\,)\big]_{+-}
&\simeq\big[\rho_{k,0}(\,\tilde t\,)\big]_{+-}+\big[T\mathcal{X}_B\,\rho_{k,0}(T_B)\mathcal{X}^{-1}_BT^\dag\big]_{+-} \e^{-\,\tilde t\,/\tau}\Theta(\,\tilde t\,-t_g)\\
&~~~~~~~~~~~~~~
+\bigg\{\big[\mathcal{X}_B\,\rho_{k,0}(T_B)\mathcal{X}^{-1}_B\big]_{+-}\e^{-\,\tilde t\,/\tau}
+\big[\mathcal{X}_BT\mathcal{X}_B\,\rho_{k,0}(T_B)\mathcal{X}^{-1}_BT^\dag\mathcal{X}^{-1}_B\big]_{+-}\e^{-(\,\tilde t\,+T_B)/\tau}\bigg\}\Theta(t_g-\,\tilde t\,),\\
&\simeq -\sqrt{P}\e^{i\delta\theta_k(t_g)-i\phi_S}\Big(1+\e^{-i(\Theta_B+2\phi_S)-T_B/\tau}\Big)\,\delta F_k(t_g)\Big\{\e^{-(\,\tilde t\,-t_g)/\tau}\Theta(\,\tilde t\,-t_g)+\e^{-i\Theta_B-(\,\tilde t\,+T_B-t_g)/\tau}\Theta(t_g-\,\tilde t\,)\Big\}\\
&~~~~~~~~~~~~~~~~~~~~~~~~~~~~~~~~~~~~~~~~~~~~~~~~~~~~~~~~~~~~~~~~~~~~~~~~~~~~~~~~~~~~~~~~~~
+\mathcal{O}(P^{3/2})+\bigg(\mbox{the terms involving $\delta F_k(t_0=0)$}\bigg).
\end{split}
\end{equation}

\endwidetext

\newpage
\bibliographystyle{apsrev4-2}
\bibliography{reference}

\begin{thebibliography}{55}
\bibitem [{Com()}]{Comment1}%
  \BibitemOpen
  \href@noop {} {\bibinfo {title} {This is exactly the same as what was
  previously referred to as the snapshot basis}},\ \bibinfo {howpublished} {in
  Ref.~\cite{PhysRevB.109.L180302}}\BibitemShut {NoStop}%
\end{thebibliography}

\end{document}